\documentclass[twocolumn,showpacs,floatfix,prl]{revtex4}

\usepackage{graphicx}
\usepackage{dcolumn}
\usepackage{bm}
\usepackage{amsmath}
\usepackage{amssymb}

\usepackage{ulem}

\begin{document}

\title{Crystallization of trions in SU(3) cold atom gases trapped in
optical lattices}

\author{Rafael A. Molina$^{1,2}$}
\author{Jorge Dukelsky$^1$}
\author{Peter Schmitteckert$^{3}$}

\affiliation{$^1$Instituto de Estructura de la Materia - CSIC, Serrano 123,
28006 Madrid, Spain \\
$^2$Institut f\"ur Theorie der Kondensierten Materie,
Universit\"at Karlsruhe, 76128 Karlsruhe, Germany \\
$^3$Institut f\"ur Nanotechnologie, Forschungzentrum Karlsruhe, Karlsruhe, Germany}

\begin{abstract}
Cold Fermionic atoms with three different hyperfine states confined in optical lattices show pronounced Atomic
Density Waves (ADWs). These ADWs are pinned due to the confining potential that traps the atoms in the optical
lattice and can be considered a crystal of strongly bound trions. We show that the crystalline phase is
incompressible and robust against SU(3) symmetry breaking interaction. We also show that it is generic in the
presence of the trap due to its incompressible characteristics. A simulation of the time evolution of the
fermionic gas after the trapping potential is switch off shows that the formation of the trionic crystal is marked
by a freezing of the dynamics in the lattice.
\end{abstract}


\maketitle


Ultracold Bose or Fermi gases confined in artificial optical lattices have become an important instrument for
investigating the physics of strong correlations. The parameters and dimensionality of these systems can be tuned
with very high precision and unprecedent control \cite{review} by means of an atomic Feshbach resonance or in an
optical lattice by changing the depth of the wells. One of the main achievements in this subject is the
experimental observation of a superfluid to Mott insulator transition in a three-dimensional optical lattice with
bosonic $^{87}Rb$ atoms \cite{Greiner02}. Very interesting experimental results have also been obtained for
fermions \cite{Stoferle06}.

High spin fermions can be directly studied with cold atoms in more than two hyperfine states. This kind of systems
could give rise to new exotic phases in optically trapped cold atoms. The general structure of high-spin Cooper
pairs has already been theoretically addressed \cite{Ho99}. More recently, the emergence of triplets and quartets
(three or four fermion bound states) has been explored \cite{Wu05,Lecheminant05,Kam05,Rapp07,Roux08}. In the case of
SU(3), fermions with three hyperfine (color) states, these investigations have suggested that it would be possible
to observe the transition from baryonic matter at low densities to a color superconducting state at higher
densities \cite{Hofstetter04,Rapp07}.
At least two alkali atoms $^6Li$ and $^{40}K$
seem to be possible candidates for the experimental realization of a SU(3)
fermionic lattice with attractive interactions\cite{Rapp07}.

The paradigmatic model used for exploring these exotic features is the one-dimensional N-component fermionic
Hubbard model with a local attractive interaction \cite{Hofstetter04}. A low-energy effective theory can be
derived through linearization of the dispersion relation like in the usual SU(2) Hubbard Hamiltonian
\cite{Giamarchibook,Assaraf99}. This theory predicts new scattering channels in the half-filled case, {\it i.e.}
the Umklapp scattering terms that do not conserve momentum but conserve quasi-momentum. In a recent work by
Capponi {\it et al.} \cite{Capponi} the bosonization parameters of SU(N) theories have been investigated by means
of large scale DMRG calculations. Regarding SU(3) their conclusion was that, for low densities, the dominant
instability is that of a trionic superfluid and that color superconductivity is strongly suppressed. There is a
competition between the superfluid trionic behavior and ADW order with the former phase dominating for generic
fillings, and the latter probably appearing at half-filling.



The main goal of this letter is to study the emergence of a trionic crystalline phase in three color atoms loaded
in optical lattices. We will show how the harmonic confinement potential can pin down an ADW at high densities in
optical lattices, that is the precursor of crystal phase of trions (CPT). As the system gains energy due to local
Umklapp scattering, the confined fermionic cloud deforms itself in the lattice to be at half-filling in the center
of the trap. Similar effects have also been reported in the repulsive and attractive SU(2) cases
\cite{Rigol,Molina07}, however for three color systems the appearance of an ADW is enhanced dramatically. The
critical value of the attractive interaction $U$ for the CPT to be fully formed is approximately $U=-4$. Two main
properties of this phase appear: a) it is an incompressible state, and b) the dynamics is almost frozen due to the
suppression of the hopping of the trions between lattice wells. Fermionic color pairs are only dominant at weak
coupling (a small region around $U=-1$).


The low energy physics of cold fermionic atoms with three different hyperfine states trapped in an optical lattice
can be described by an SU(3) generalization of the Hubbard Hamiltonian,
\begin{eqnarray}
H&=&-t\sum_{\left\langle ij\right\rangle \alpha} \left(f_{i\alpha}^{\dagger}f_{j\alpha }+{\mathrm{h.c.}}\right)+ \nonumber
\sum_{i\alpha\neq\beta} \frac{U_{\alpha\beta}}{2} n_{i\alpha}n_{i\beta} \\
&+&V\sum_{i\alpha}(i-L/2)^2~n_{i\alpha}.
\label{eq:Hamiltonian}
\end{eqnarray}

The sums over $\alpha$ and $\beta$ go over the three colors
The operators
$f_{i\alpha}^{\dagger}$ and $f_{i\alpha }$ are the creation and destruction operators of an atom in site $i$ with
color $\alpha$. We consider different values of the on-site interaction between the different color pairs
$U_{\alpha\beta}$ to be able to include SU(3) symmetry breaking terms as they would probably be present in future
experiments. The site label $i$ goes from $0$ to $L-1$, with $L$ being the total number of lattice sites. All
energies are expressed in units of $t$ ($t=1$).




We study the ground state properties of the Hamiltonian (\ref{eq:Hamiltonian}) with the DMRG algorithm \cite{DMRG}
that provides very accurate numerical results. The modifications of the DMRG procedure due to the trap potential
that breaks the translational symmetry have been described in previous publications
\cite{schmitteckert,schmitteckert_werner,Molina07}. In order to obtain enough accuracy we have kept up to 1200
states in each iteration.

\begin{figure}
\includegraphics[width=8cm]{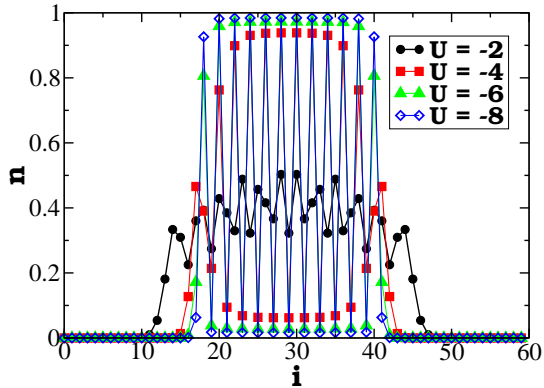}
\caption{(Color online)
Density for different values of $U$ for $L=60$, $N=36$, and $V=0.001333$.}
\label{fig:l60q36d0}
\end{figure}

In Figure \ref{fig:l60q36d0} we show the atom density for equal color populations $N=N_r+N_g+N_b=36$, parameters
$V=0.003$ , $L=60$, and different values of the attractive SU(3) interaction $U=U_{rg}=U_{rb}=U_{gb}$. The figure
clearly shows the formation of the CPT as $|U|$ increases. For $U \leq -4$ the phase is fully developed.
Increasing $|U|$ compresses the density as interaction energy gains over kinetic energy and the particles prefer
to be in the center of the trap. However, the crystal phase is incompressible and, once it is formed, increasing
the value of the attraction does not compress the particle density any further.

The formation of these phases can be studied by means of local correlation functions that measure the number of
pairs and the number of trions. For the number of pairs we define
\begin{equation}
P_{i \alpha \beta}=\left< n_{i \alpha} n_{i \beta} \right>-
\left< n_{ir} n_{ig} n_{ib}\right>.
\label{eq:pairs}
\end{equation}
In the non-polarized we will be interested in the total number of pairs at each site i,
\begin{equation}
P_i=P_{irg}+P_{irb}+P_{igb}.
\label{eq:total_pairs}
\end{equation}
The local correlation that counts the number of trions is
\begin{equation}
T_i=\left< n_{ir} n_{ig} n_{ib}\right>.
\end{equation}
For non-interacting SU(3) fermions
$P_{i\alpha\beta}=\left< n_{i\alpha}\right> \left< n_{i\beta}\right>-\left< n_{ir} \right> \left< n_{ig} \right>\left < n_{ib}\right>$ and
$T_i=\left< n_{ir} \right> \left< n_{ig} \right>\left < n_{ib}\right>$.

These local correlations can be summed over all the sites on the system to obtain the total number of fermions in
pairs (2P) and the total number of fermions in trions (3T). The number of uncorrelated fermions (F) is then
$F=Q-2P-3T$.



\begin{figure}
\includegraphics[width=8cm]{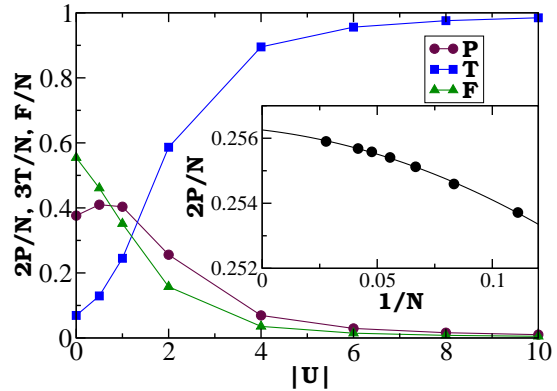}
\caption{(color online) Extrapolated total number of fermions in pairs (2P) and in trions(3T) compared with the
single fermions(F) for different sizes as a function of $|U|$. Inset: Fitting second order polynomial for $2P/N$
as a function of $1/N$ for $U=-2$.} \label{fig:extrapolated_correlations}
\end{figure}

The proper finite size scaling of the systems, as described in \cite{Damle}, is given by $N^2V=constant$ when $N
\rightarrow \infty$ and $V \rightarrow 0$. We checked the validity this scaling law by comparing results for
different sizes. For a given value of $U$ the results for the fraction of fermions in trions $3T/N$ and the
fraction of fermions in pairs $2P/N$ as a function of the inverse size $1/N$ can be easily fitted by a second
order polynomial. In this way we can extrapolate the values of trion and pair fractions in the thermodynamic
limit. Figure \ref{fig:extrapolated_correlations} shows these results. It can be seen that the number of trions
increases rapidly with the interaction strength until almost all fermions are present as trions for $U<-4$. The
number of pairs is only dominant in a small region around $U=-1$ suggesting a color superconducting phase. An
example of the quality of the extrapolation procedure is shown in the inset of the figure.

To understand better the origin of the ADW state we will compare the DMRG ground state energy with several
approximations. In the strong interaction limit it is possible to treat the hopping term as a perturbation.
Ignoring the effect of the trap, the zeroth order energy of each trion is $3U$. The first non-zero correction
appears in second order perturbation theory giving a lowering of $3t^2/U$ for two next to nearest-neighbor trions,
due to quantum fluctuations of the fermions within the trions. If two trions sit in neighboring sites this term is
completely suppressed. Therefore, it acts as an effective repulsion between nearest-neighbor trions, being at the
origin of the CPT at half-filling \cite{Zhao07}. Unlike the SU(2) case, the first contribution to hopping appears
in the third order of perturbations, thus a CPT will always dominate for strong enough attraction.

\begin{figure}
\includegraphics[width=8cm]{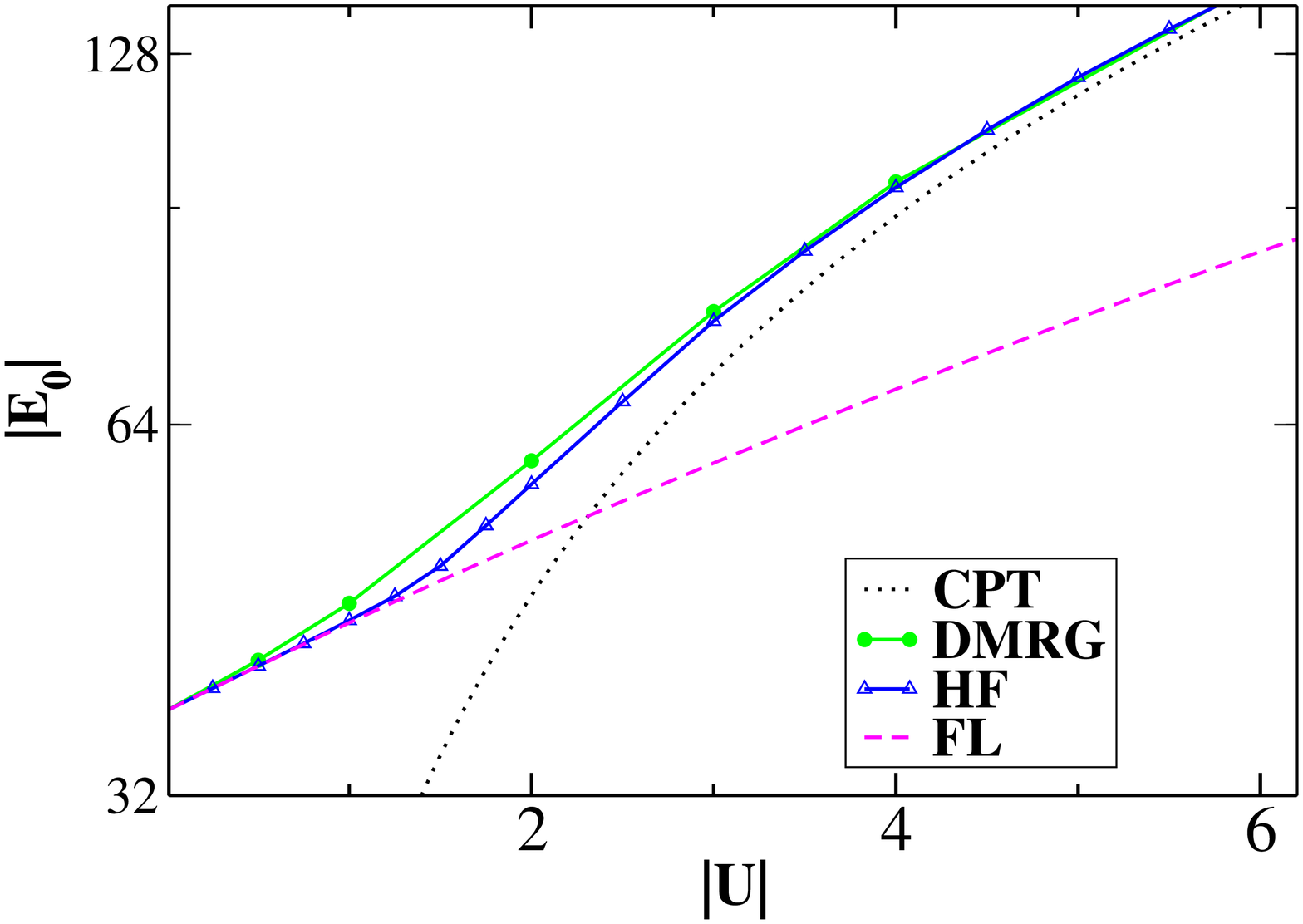}
\caption{(Color online) Absolute value of the ground state energy $|E_0|$ in logarithmic scale as a function of
$|U|$ for the case with $L=40$, $V=0.003$, and $N=24$. DMRG results are compared with a HF approximation, a CPT,
and a Fermi liquid state of trapped trions.} \label{fig:HF}
\end{figure}

Adding the effect of the trap we have calculated the energy of the CPT state, defined as
$\prod_{i'}c^{\dagger}_{i',r} c^{\dagger}_{i',g} c^{\dagger}_{i',b} \left|0\right>$, where the $i'$ labels the
even sites closer to the trap center until all the fermions are accounted for. We have also calculated the energy
of a Fermi liquid (FL) state, defined as a Slater determinant in the basis that diagonalizes the hopping term plus
the trapping potential. The CPT state valid in the strong coupling limit and the FL state appropriate for weak
coupling limit are both Slater determinants independent of the attractive interaction $U$. Therefore, it is
expected that a Hartree-Fock (HF) approximation that looks for the best Slater determinant for a given value of
$U$ will interpolate between the two limits, and it will provide a better approximation for intermediate
interactions. In Fig. \ref{fig:HF} the ground state energy $E_0$ as a function of $|U|$ for the three different
approximations is compared with the DMRG results. As expected, the HF energies are very accurate for small and
large values of $|U|$. They deviate from the DMRG results appreciably in a region around $U=-2$ where the
intermediate color superconductivity phase appears to take place.

\begin{figure}
\includegraphics[width=8cm]{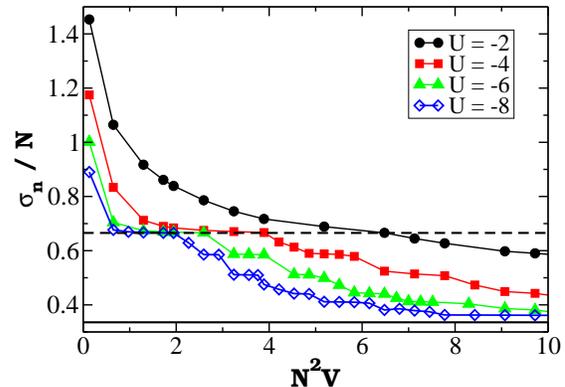}
\caption{(Color online) Normalized size of the fermionic cloud $\sigma_n/N$
as a function of $N^2V$ for different values of $U$ and $N=36$.
}
\label{fig:compres}
\end{figure}

When the effect of the trap is not taken into account the ADW order is non-generic occurring only at exactly half-filling.
According to our simulations the CPT state {\it is} generic in the presence of a trap and occurs for a wide range
of values of the number of particles in the trap, the confining potential, and the interaction strength. To address this issue we have calculated the
extension of the fermionic cloud in the trap, defined as $\sigma_n=\sqrt{\sum_i n_i(i-L/2)^2}$, as we vary the
confining potential for different values of the interaction strength. Results of these calculations for $N=36$ are
shown in Fig. \ref{fig:compres}. The horizontal dotted line denotes the value of $\sigma_n$ for the CPT while the
continuous horizontal line at the bottom denotes the Band Insulator (BI) state with all atoms filling the bottom
of the trap. For $U \leq -4$, when the value of the CPT is reached, there is a plateau of $\sigma_n$ as a function
of $N^2V$ marking the appearance of an incompressible state. For $U \leq -6$, new plateaus for lower $\sigma_n$
and higher values of $V$ appear. These other plateaus correspond to the presence of a band insulator in the center
and symmetric CPTs at the sides of the trap. The crossover from the CPT to the BI as we increase the confinement
consists on a number of steps in which each of the trions in the CPTs in the outer side of the cloud is forced to
move the BI in the center (an analogous phenomenom has been recently described for spin 3/2 atoms \cite {Roux08}).

\begin{figure}
\includegraphics[width=4cm,height=4cm]{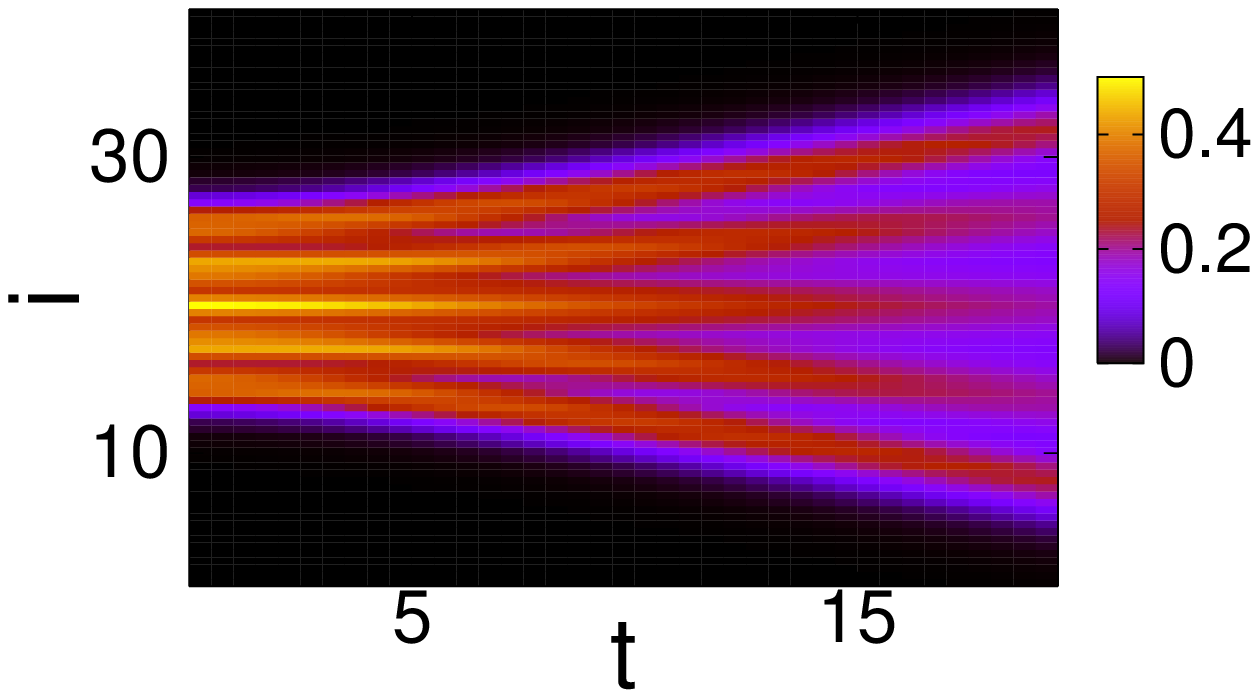}
\includegraphics[width=4cm,height=4cm]{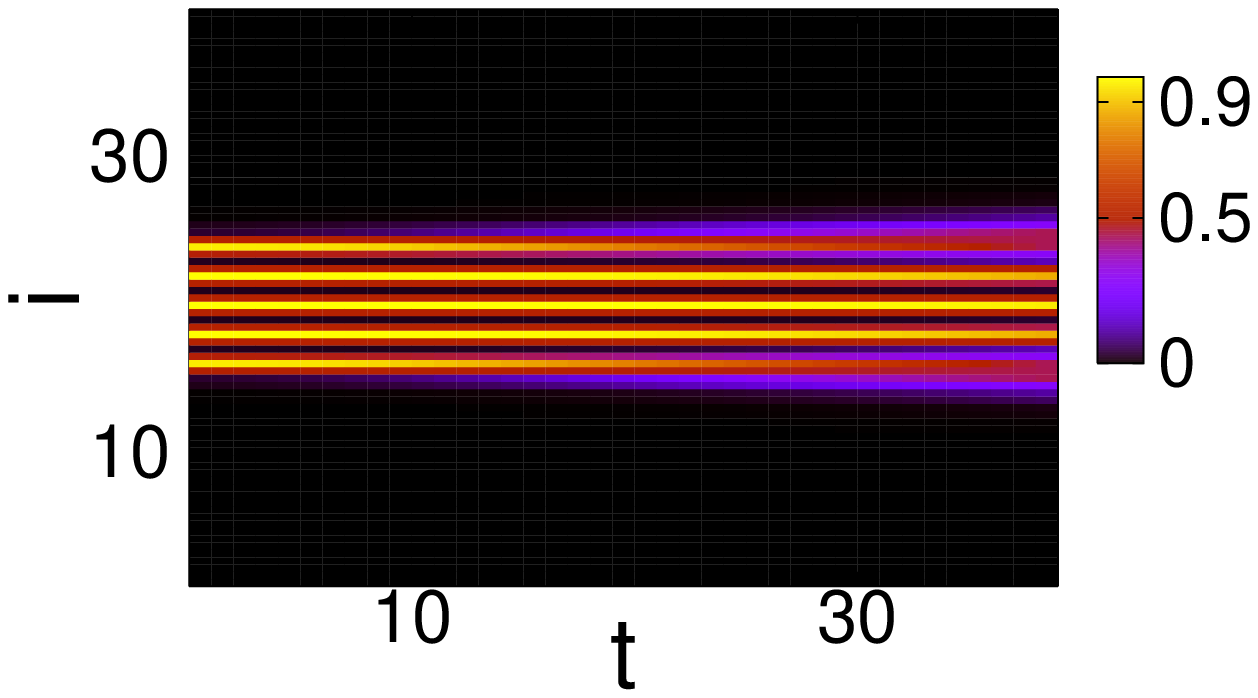}
\caption{(Color online) Evolution of the ground state density after a sudden switch off of the confinement
potential for the case with $L=40$, $V=0.005$, and $Q=15$. Left panel: $U=-2$. Right panel: $U=-8$.
}
\label{fig:time1}
\end{figure}

\begin{figure}
\includegraphics[width=8cm]{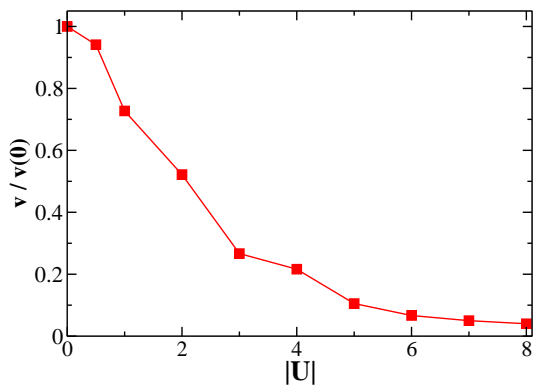}
\caption{(Color online)
Velocity of expansion as a function of the interaction in a case with $N=15$.}
\label{fig:vel}
\end{figure}

In what follows we will explore the evolution of the CPT when the trap is opened. We expect that its dynamics
 should be frozen since the effective hopping of a bound trion is extremely small (third order in perturbation theory).
We have prepared a state of five trions made out of 15 fermions in a lattice of size $L=40$ and $v=0.005$. The
harmonic confinement potential is switched off at $t=0$, and the state evolves with the rest of the Hamiltonian
(\ref{eq:Hamiltonian}). The temporal evolution is computed with the time dependent DMRG algorithm \cite{Petert}.
In Fig. \ref{fig:time1} we show the evolution in natural time units of the density for two selected values of $U$:
 $U=-2$ in the color superconducting phase and $U=-8$ well in the CPT. As a general trend, we observe that the
expansion of the cloud almost ceases for $U\leq -6$. In order to quantify this statement we have calculated the velocity
of the most external trion as a function of $U$. The velocity is computed by means of a linear fit of the position
of the external peak as a function of time. The results are depicted in Fig. \ref{fig:vel} where can see how the
velocity is strongly suppressed for $U<-4$.



As already mentioned, a possible candidate for realizing the trionic phase is $^6Li$ for which the magnetic field
dependence of the three scattering lengths has been measured \cite{exp6li}. The attractive $U$ interaction in some
appropriate regions of magnetic field can be estimated as $U_{rg}=U_0$, $U_{rb}=1.23U_0$, and $U_{gb}=1.06U_0$
\cite{Rapp07}. Furthermore, the effective value of $U_0$ can be controlled changing the depth of the optical
lattice wells. We have numerically checked the effects of this experimental anisotropy in the
interaction. The CPT persists without noticeable changes even for stronger SU(3) symmetry breaking interactions.

In summary, we have shown that an CPT of trions is a very robust and generic phase of atoms with three hyperfine
states confined in optical lattices. Due to the incompressible properties of this state it should be easily
observed for experiments with high-density of fermionic atoms. The appearance of this state is marked by a
freezing of the size of the cloud when the trapping  potential is switched off, as we have shown through a full
time simulations of the evolution. In future experiments of ultracold fermionic atoms the existence of the CPT
could be revealed by light-scattering diffraction experiments. Moreover, radio-frequency spectroscopy could be
used to probe the dissociation dynamics of the trions.


The authors acknowledge discussions with G. Zarand, P.
 Lecheminant, and
R. Thomale. This work
is supported in part by Spanish Government grant
No. FIS2006-12783-C03-01 and by grant CAM-CSIC
No. CCG07-CSIC/ESP-1962. RAM contract is financed by CSIC and
the European Comission through the I3p program and by a Jose Castillejo
Grant of the Spanish Ministry of Research.

\end{document}